\documentclass[12pt]{article}

\usepackage{amssymb}
\usepackage{amsmath}
\usepackage{mathrsfs}

\def\ba{\begin{eqnarray}}

\def\ea{\end{eqnarray}}
\def\14{{1\over4}}
\def\12{{1 \over 2}}
\def\ap{area preserving diffeomorphism}

\def\ep{\epsilon_{ij}}

\def\ro{\rho_0}
\def\epa{\epsilon_{ab}}

\def\ncy{non-commutativity}

\def\ah{\hat{A}}

\newcommand{\be}{\begin{equation}}
\newcommand{\bea}{\begin{eqnarray}}
\newcommand{\eea}{\end{eqnarray}}
\newcommand{\beq}{\begin{equation}}
\newcommand{\ee}{\end{equation}}
\newcommand{\eeq}{\end{equation}}

\begin{document}

\vspace{10mm}
\begin{center}
{\Large  \bf Hierarchy Construction of Quantum Hall States and
Non-Commutative Chern-Simons Theory }

\vspace{10mm} {\large Zhao-Long Wang\footnote{E-mail address:
zlwang4@mail.ustc.edu.cn}, Wei Huang\footnote{E-mail address:
weihuang@mail.ustc.edu.cn}, and Mu-Lin Yan\footnote{E-mail address:
mlyan@ustc.edu.cn}\\} \vspace{10mm}
 {\it
Interdisciplinary Center for Theoretical Study, Department of Modern Physics\\
University of Science and Technology of China, Hefei, Anhui 230026,
China}
\end{center}

\begin{abstract}
In this paper, we study the non-commutative Chern-Simons description
of the hierarchy of quantum Hall states. Our method is based on the
framework suggested by Susskind in hep-th/0101029. By using the area
preserving diffeomorphism gauge symmetry of quasiparticle fluid, we
show that non-commutative Chern-Simons description of the hierarchy
construction of quantum Hall states with generic filling fraction
can be realized in Susskind's approach. The relationship between our
model and the pervious work on the effective field theory of quantum
Hall states is also discussed.
\end{abstract}

\section{Introduction}

After the discovery of fractional Quantum Hall Effect (FQHE) in
1982, many efforts have been made on the theoretical explanation of
this phenomenon. As pointed in \cite{FQHE} (see also
\cite{laghl1}\cite{laghl2}), the most well established theoretical
idea in the domain of the FQHE is the Laughlin wave function
describing the states with filling fraction $\nu={1\over 2m+1}$,
\be\label{1} \Psi_{laugh} = \prod_{i<j} (Z_i-Z_j)^{1\over \nu} \exp{
\left( -\12 \sum |Z_i|^2 \right)} .\ee The main developments on the
states with more generic filling fraction are the hierarchy
construction \cite{haldane} as well as the composite fermion
proposal \cite{jain}. In the hierarchy construction, the
condensation of quasiparticles changes the filling fraction of the
system naturally. Any states with odd denominator filling fraction
could be realized in this manner. On the other hand, the composite
fermion proposal suggests that electron combining with 2p magnetic
flux quanta forms a composite fermion. The FQHE is explained as the
integer Quantum Hall Effect of the composite fermion. The composite
fermion approach have made great progress on explaining the
experiment results. However, the physical mechanism of forming the
composite fermion remains a great puzzle.

Many effective field theories based on these two approaches were
established (see for example \cite{wen}\cite{lopez}\cite{scz}). An
interesting observation is that the Chern-Simons theory plays a
crucial role in both the two approaches. For example, the
Chern-Simons term provides the mechanism of transition between
electron and composite fermion, and the coefficients in the
effective Chern-Simons action of hierarchy model is suggested to
describe the topological order of the corresponding state. Despite
the successes of these effective on describing the physics of FQHE,
the origin of the Chern-Simons term remains unclear. A reasonable
idea on this item in which the Chern-Simons action arises naturally
from the Lorentz force of magnetic field is suggested in
\cite{Bahcall} by Bahcall and Susskind. Inspired by the development
of matrix model in M theory, this idea is developed more
systemically and more soundly in \cite{susskind}: 1) The dynamics of
2D-electron fluid in perpendicular external strong magnetic field is
a Chern-Simons gauge theory based the group of area preserving
diffeomorphisms (APD's). This gauge theory of APD's captures many of
the long distance features of the Quantum Hall system; 2) In order
to capture the discrete or granular character of the electron
system, we need to discretize the APD's. A well known way to do so
is method of the non-commutative field theory. Thus, a
non-commutative version of Chern-Simons theory for FQHE is achieved.
This scenario can be thought of as a micro-theory to describe FQHE
based on the dynamics of noncommutative fluids of charged particle
(e.g., see a recent review \cite{poly}). The corresponding matrix
model has also been established. The subsequent discussions
\cite{polychronakos}\cite{hellerman}\cite{karabali}\cite{cappelli}
have shown the consistence between this description and the Laughlin
wave function's with $\nu={1\over 2m+1}$. Following this great
success, we should pursue a further challenge problem whether the
hierarchy construction of $\nu$ could be derived from this fluid
dynamics theory or not. In other words, we should pursue the micro
dynamical origin of the hierarchy construction in FQHE. The purpose
of this paper is reformulating the hierarchy construction in the
Susskind's framework. The main point is the condensation of
quasiparticles which are the excitations of original Quantum Hall
Fluid could also be analyzed in the Susskind's approach to get a
non-commutative Chern-Simons description.

The content is organized as follows. We review the argument of
\cite{susskind} first in Section 2. Then filling fraction hierarchy
is constructed by considering the condensation of quasiparticles in
Section 3. In Section 4, the effects of interaction between
quasiparticles is discussed. we show that these interaction is
related to the fractional statistics of quasiparticles. The full
noncommutative Chern-Simons theory as well as the corresponding
matrix model is established. Finally, the relationship with previous
results is discussed in Section 5.

\section{Non-commutative Chern-Simons description of quantum Hall effect}

In order to illuminate the clue we will follow and the notations, we
briefly recall the procedure in \cite{susskind}. A collection of
identical electrons indexed by $\alpha$, moving on a plane can be
described by the Lagrangian \be\label{2} L=\sum_{\alpha}{m\over
2}\dot{x}_{\alpha}^2 -V(x) ,\ee where $V$ is the potential energy.
Assuming the system behaves like a fluid we can pass to a continuum
description by replacing the discrete label $\alpha$ by a pair of
continuous coordinates $y_1,y_2$. These coordinates label the
material points of the fluid. The system of particles is thereby
replaced by a pair of continuum fields $x_i(y,t)$ with $i=1,2$.
Without loss of generality we can choose the coordinates $y$ so that
the number of particles per unit area in $y$ space is constant and
given by $\ro$. The real space density is \be\label{3} \rho = \ro
\left| {\partial y \over
\partial x}\right| ,\ee where $ \left| {\partial y \over \partial
x}\right|$ is the Jacobian connecting the $x$ and $y$ coordinate
systems. One further assume that under this fluid approximation the
potential terms could be written as \be\label{4} V=V\left(\ro\left|
{\partial y \over \partial x}\right|\right) .\ee The Lagrangian
becomes \be\label{5} L=\int d^2 y \ro \left[ {m\over 2} \dot{x}^2 -
V\left(\ro \left| {\partial y \over \partial x}\right|\right)
\right] .\ee The fluid is dissipationless and the above Lagrangian
has a gauge symmetry originated from relabeling the particles.
Consider the \ap \ from $y$ to $y'$ with unit Jacobian. It induce
the fields transformation $x'(y') =x(y)$ which keeps the Lagrangian
invariant if the boundary terms has no contribute (e.g. the
periodical boundary condition). The infinitesimal transformation
takes the form \be\label{6} \delta x_a= \ep {\partial
x_a\over\partial y_i} {\partial \Lambda \over
\partial y_j} .\ee

Under the assumption that the potential $V$ in (\ref{4}) has a
minimum at $\rho = \ro$, \cite{susskind} defined a field $A$ by
\be\label{7} x_i = y_i +\ep {A_j \over 2\pi \ro} .\ee The gauge
transformation becomes \be\label{8} \delta A_i = 2\pi \ro {\partial
\Lambda \over \partial y_i} +{\partial A_i \over \partial
y_l}{\partial \Lambda \over \partial y_m} \epsilon_{l,m} .\ee which
has the form as the first order truncations of a U(1) gauge
transformation on non-commutative plane. Thus approximately $A$
could be regarded as a U(1) gauge field on the non-commutative
$y$-plane. Further discussion shows that (\ref{5}) will give the
standard Maxwell action of $A$ under proper gauge conditions.

If there is a background magnetic field $\mathcal{B}$, the
Lagrangian gets an extra term \be\label{9} L'= {e{\mathcal{B}} \over
2} \int \ro d^2y \epa \dot{x}_a x_b .\ee Substituting (\ref{7}) into
(\ref{9}) and dropping total time derivatives gives the desired
Chern-Simons (CS) Lagrangian \be\label{10} L'= {e{\mathcal{B}} \over
8\pi^2 \ro} \int d^2y \epa \dot{A}_a A_b=\frac{1}{4\pi\nu_1}\int
d^2y \epa \dot{A}_a A_b .\ee If $\mathcal{B}$ is sufficiently large,
the dynamics will be dominated by CS term and the Maxwell terms
could be ignored.

For (\ref{9}), the conserved quantity related to the gauge symmetry
is \be\label{11} qJ^0= \frac{e{\mathcal{B}}}{2\pi} \left(\12 \ep
\epa {\partial x_b \over \partial y_j} {\partial x_a \over \partial
y_i}-1\right) ,\ee where $q$ is charge of the quasiparticles, that
is, the vortices of fluid. To linear level, (\ref{11}) becomes
\be\label{12} \frac{1}{2\pi\nu_1} \nabla \times A = q J^0(y) .\ee
Take $ J^0=\delta^2(y)$, one can derive the quantization of the
charge $q$ as in \cite{susskind} $$ q=n ,$$ where $n$ is integer.

Projected back to the $x$-plane, it is shown in \cite{susskind} that
the quasiparticle carries fractional electronic change
$e_{qp}=q\nu_1 e$ as well as magnetic flux $\Phi_{qp}=q (2\pi/ e)$.
The statistic phase for exchanging two quasiparticles is
$\phi_{qp}=\pi\nu_1$, that is, the quasiparticle obeys the
fractional statistics.

If $q=0$ there is no vortices excitation, one can incorporate the
constraint to the lagrangian by introducing a time component of $A$:
\be\label{14} L'={e \mathcal{B} \ro \over 2} \epa \int d^2y \left[
 \left(\dot{x_a} - {1\over 2 \pi \ro}\{x_a,A_0
\}\right)x_b -{\epa \over 2\pi \ro}A_0 \right ] .\ee where $ \{
F(y),G(y) \} = \ep \partial_iF
\partial_jG
$ denotes the Poisson bracket. Furthermore, because of the
observation that (\ref{8}) and(\ref{14}) have the form as the first
order truncations of a U(1) CS theory on non-commutative plane, as
well as considerations about the discreteness of electrons, it is
argued in \cite{susskind} that the theory describing the quantum
Hall system should be the non-commutative CS theory \be\label{15}
L=-{1 \over 4 \pi \nu_1} \int d^2y\epsilon^{\mu \nu \rho} \left(
\ah_{\mu}\ast
\partial_{\nu} \ah_{\rho} +{2i \over 3} \ah_{\mu} \ast \ah_{\nu}
\ast \ah_{\rho} \right) ,\ee with the usual Moyal star product
defined in terms of the \ncy \ parameter \be\label{16}
[y_1,y_2]=i\theta_1 = {i \over 2\pi \ro} ,\ee and
$$\nu_1={2\pi\over e\mathcal{B}}\rho_0 .$$ This non-commutative CS
theory can be reformulated via the following matrix model,

 \be\label{17} L_M={e\mathcal{B}\over 2}\epa Tr \left( \dot{x_a}+i [x_a,
A_0]_m \right)x_b -e\mathcal{B} \theta Tr A_0 .\ee where $x_a$ and
$A_0$ are infinite dimension hermitian matrices. By considering the
operation of exchange two particles in the matrix model, one can
find the important result that the statistic phase of the particles
which constitute the fluid is related to the filling fraction as
\be\label{18} \phi={\pi\over \nu_1} .\ee Thus, for the electron
fluid the level parameter in (\ref{15}) should be quantized
as\be\label{19} \nu_1=\frac{1}{p_1} ,\ee where $p_1$ is an odd
integer.

In absence of the quasiparticle excitations, the static solution
$x_i=y_i$ implies that the electron density on the physical
$x$-plane is $ \rho=\ro$ . Therefore, one gets an explanation of the
filling fraction of the fractional quantum Hall effect
$$\nu=\frac{2\pi\rho}{e{\mathcal{B}}}={1\over p_1} .$$

\section{The quasiparticle condensation and the hierarchy construction}

In order to describe the quantum Hall effects with generic filling
fraction, we should consider the condense of quasiparticles on the
quantum Hall fluid.

First consider the dynamics of one quasiparticle. According to the
quantization condition of $q$, for the element quasiparticle, we
have $q=\pm 1$. Regarding the quasiparticle as ``electron'' on the
$y$-plane, and now the $A_i$ plays the role of electromagnetic
field. Analog with the electromagnetic coupling, one may take the
form as
$$ L_{int}=q\int d^2y A_{\mu}J^{\mu}+...$$
The dots refer to possible higher order correction from the
non-commutativity of $y$-plane. We will come back to this issue in
the next section. For a single quasiparticle, the background value
of $A_i$ is zero, therefore this term vanishes.

On the other hand, in the $x$-plane viewpoint the quasiparticle
provides an extra electric charge $ e_{qp}=q\nu e$. The
electromagnetic interaction for this extra charge is not taken
account in the origin lagrangian (\ref{14}). Noting that the
background configuration for the quasiparticle is
$$ x_i=y_i ,$$ the corresponding interacting term should be
\be L'_{int}=-\frac {e_{qp}\mathcal{B}}{2}\epa \dot{X}_{a} X_{b}
=-q\frac {2\pi\ro}{2}\epa \dot{Y}_{a} Y_{b} ,\ee where $X$ and $Y$
are the position of quasiparticle on $x$-plane as well as $y$-plane.
The magnetic field $\mathcal{B}$ on $x$-plane has induced a
effective ``magnetic field'' on $y$-plane \be\label{20} B=2\pi\ro
.\ee In presence of this background magnetic field $B$, the
coordinate operator will get a non-commutativity
$$[\hat y_1,\hat y_2]=-\frac{i}{q B}=-\frac{i}{2\pi\ro q} .$$
It is consistent with the the non-commutativity of $y$-plane
$$[y_1,y_2]=\frac{i}{2\pi\ro} .$$

Now if there is a quasiparticle fluid on the $y$-plane, one can
repeat the procedure in Section 2. The quasiparticles are labeled by
a pair of continuous coordinates $z_i$, and the position of
quasiparticle on the $y$-plane becomes the continuum fields
$y_i(z,t)$. Without loss of generality we can choose the coordinates
$y$ so that the number of particles per unit area in $y$ space is
constant and given by $\ro'$. Furthermore, one assumes that the
potential energy from the Coulomb interaction becomes a function of
density under this fluid approximation, that is, analog of (\ref{4})
holds for the quasiparticles. Therefore, a gauge symmetry originated
from relabeling holds in this description. The corresponding
infinitesimal transformation is \be\label{21} \delta y_a= \ep
{\partial y_a\over\partial z_i} {\partial \Lambda' \over
\partial z_j} .\ee Under the assumption that $\rho '=\ro '$ is a
local minimum, one defines \be\label{22} y_i = z_i +\ep {a_j \over
2\pi \ro'} ,\ee and the gauge transformation becomes\be\label{23}
\delta a_i = 2\pi \ro' {\partial \Lambda' \over \partial z_i}
+{\partial a_i \over
\partial z_l}{\partial \Lambda' \over \partial z_m} \epsilon_{l,m} .\ee
This suggests that the theory describing the small deviation from
the minimum should be a U(1) gauge theory on the noncommutative
$z$-plane.

If we ignore kinetic term as well as the interaction between the
quasiparticles, we could get the second level non-commutative CS
action following the same argument as that in Section 2
\be\label{24} L_2=-{q \over 4 \pi \nu_2} \int d^2z\epsilon_{\mu \nu
\rho} \left( a_{\mu}\star
\partial_{\nu} a_{\rho} +{2i \over 3} a_{\mu} \star a_{\nu}
\star a_{\rho} \right) ,\ee with the usual Moyal star product
defined in terms of the \ncy \ parameter \be\label{25}
[z_1,z_2]=i\theta_2 = {i\over 2\pi \ro'} .\ee The level parameter
$\nu_2$ is defined as \be\label{26}
\nu_2=\frac{2\pi\ro'}{B}=\frac{\ro'}{\ro} .\ee Since we ignore the
interaction between quasiparticles, the quasiparticle should be
regarded as an boson on the $y$-plane. In order to give the right
statistics in the corresponding matrix model, the level parameter in
(\ref{24}) should be quantized as \be\label{27} \nu_2=\frac{1}{p_2}
,\ee where $p_2$ is a even positive integer since both $\ro'$ and
$\ro$ are positive.

In our present case, the constraint (\ref{11}) takes the form
\be\label{28} \frac{e{\mathcal{B}}}{2\pi}\left(\12 \ep \epa
{\partial x_b \over
\partial y_j} {\partial x_a \over \partial y_i}-1\right)=q\ro' .\ee
Then the electron density on the physical $x$-plane becomes
\bea\label{29} \rho &=& \ro\left| {\partial y \over
\partial x}\right| \cr &=& \ro\left(\12 \ep \epa {\partial x_b \over
\partial y_j} {\partial x_a \over \partial y_i}\right)^{-1} \cr
&=& \ro \left(1 + \frac{2\pi q\ro'}{e {\mathcal{B}}} \right)^{-1}
\cr &=& \ro \left(1 +q \nu_1 \nu_2 \right)^{-1} .\eea Thus, one gets
the filling fraction \bea\label{30} \nu &=&
\frac{2\pi\rho}{e{\mathcal{B}}}=\nu_1 \left(1 +q \nu_1 \nu_2
\right)^{-1}\cr &=& \frac{1}{p_1+\frac {q}{p_2}} .\eea The same
procedure could be carried on level by level, and we could
reformulate the hierarchy construction \cite{haldane} naturally in
Susskind's approach \bea\label{31} \nu
 =\frac{1} {p_1\pm\frac {1}{p_2\pm\frac {1}{p_3\pm\dots}}} .\eea
Since that $p_1$ is odd and $p_i(i>1)$ is even, any filling fraction
with odd denominator could be realized in this manner.

\section{Interaction between the quasiparticles}

Now we incorporate the interaction between the quasiparticles. There
are two kinds of interactions in our system. The first one is the
Coulomb interaction of electronic charges. This interaction could be
approximated a potential which is a function of the quasiparticle
density as assumed before. If $B$ is large, this term could be
ignored when one consider the long range behavior. The second one
comes from the $y$-plane magnetic field produced by the
quasiparticle fluid itself in (\ref{12}). The mean value of this
magnetic field is \be\label{32} B_{self}=2\pi\nu_1 q\ro' ,\ee thus
the total magnetic field felt by quasiparticle should be
\be\label{33} B_{total}=2\pi\nu_1 q\ro' +2\pi\ro ,\ee and the total
filling fraction is altered as \be\label{34} \tilde\nu_2={2\pi\ro'
\over q B_{total}}={1\over q^2\nu_1+{q\over \nu_2}} .\ee Then the
corresponding matrix model has the statistic phase \be\label{35}
\Delta\Gamma={\pi \over \tilde\nu_2}=\pi\left({q^2\nu_1+{q\over
\nu_2}}\right) .\ee Since $1/\nu_2$ is even integer, the phase is
equivalent to $\pi\nu_1$ which is just the statistic phase of
quasiparticle obtained in \cite{susskind}. This result is not
surprising. In \cite{susskind}, the derivation of quasiparticle
statistics is just based on considering the back reaction of
quasiparticle.

Now we will decide the action of the full theory. As in
\cite{susskind}, the initial Lagrangian comes from the Lorentz force
in the magnetic field is \be\label{36} L=q\ro'\int
d^2z\left[\left(-{B\over 2}\epsilon_{ij} y_j+\hat A_i+{\theta_1\over
2}\epsilon_{jk}\hat A_j{\partial \hat A_k\over\partial
y_i}\right)\dot y_i -\hat A_0\right] ,\ee where \be\label{37} \hat
A_{\mu}(z)=A_{\mu}(y(z)) .\ee Noting that the field strength on the
non-commutative $y$-plane is $
F_{\mu\nu}=\partial_{\mu}A_{\nu}-\partial_{\mu}A_{\nu}-i(A_{\mu}\ast
A_{\nu}-A_{\nu}\ast A_{\mu})$, we should make sure this field
strength is felt by the quasiparticle in order to obtain the right
statistics. Therefore, additional term is added to the
$A_{\mu}j^{\mu}$ coupling. Here we just write down the linear term
appeared before we take the full non-commutative assumption as in
(\ref{15}).

The conserved quantity related to the gauge transformation
(\ref{23}) is \be\label{38} \int d^2 y \Pi_i \delta y_i ,\ee where
$\Pi_i$ is the canonical conjugate to $y_i$ \be\label{39}
\Pi_i=q\ro'\left(-{B\over 2}\epsilon_{ij} y_j+\hat
A_i+{\theta_1\over 2}\epsilon_{jk}\hat A_j{\partial \hat
A_k\over\partial y_i}\right) .\ee Then we obtain the constraint
analogy with (\ref{12}) \bea\label{40}
&&\epsilon_{kl}{\partial\over\partial z_l}\left(-{B\over
2}\epsilon_{ij}y_j+\hat A_i+{\theta_1\over 2}\epsilon_{jm}\hat
A_j{\partial \hat A_m\over\partial y_i}\right){\partial
y_i\over\partial z_k} \cr &=&-{B\over
2}\epsilon_{ij}\epsilon_{kl}{\partial y_j\over\partial z_l}{\partial
y_i\over\partial z_k}+\epsilon_{kl}{\partial \hat A_i\over\partial
z_l}{\partial y_i\over\partial z_k}+{\theta_1\over
2}\epsilon_{kl}\epsilon_{jm}{\partial \hat A_j\over\partial
z_l}{\partial \hat A_m\over\partial z_k}\cr
&=&-B\left(1+q\nu_1\nu_2\right) .\eea

Now we introduce the axillary field $a_0$ and take the constraint as
equation of motion for $a_0$. The Lagrangian becomes \bea\label{41}
L&=&q\ro'\int d^2z\left(-{B\over 2}\epsilon_{ij} y_j+\hat
A_i+{\theta_1\over 2}\epsilon_{jk}\hat A_j{\partial \hat
A_k\over\partial y_i}\right)\left(\dot y_i-{1\over
2\pi\ro'}\{y_i,a_0\}\right)\cr&&+{B\over
2\pi\ro'}\left(1+q\nu_1\nu_2\right)a_0-\hat A_0 .\eea Expanding
around the vacuum \be\label{42} y_i = z_i +\ep {a_j \over 2\pi \ro'}
,\ee one deduced the Lagrangian for the fluctuation as
\bea\label{43} L&=&-{q\over 4\pi\nu_2}\int
d^2z\epsilon_{\mu\nu\rho}\left(a_{\mu}\partial_{\nu}
a_{\rho}+{\theta_2\over 3}a_{\mu}\{a_{\nu},a_{\rho}\}\right)\cr
&&+{q\over 2\pi}\int d^2z\left(\epsilon_{ij}\left(\hat
A_i+{\theta_1\over 2}\epsilon_{lk}\hat A_l{\partial \hat
A_k\over\partial y_i}\right)f_{0j}-2\pi\ro'\hat
A_0+Bq\nu_1\nu_2a_0\right) ,\eea where \be\label{44}
f_{\mu\nu}=\partial_{\mu}a_{\nu}-\partial_{\nu}a_{\mu}+\theta_2\{a_{\mu},a_{\nu}\}
.\ee In this case, the constraint can be rewritten as \be\label{45}
{B\over4\pi\ro'}\epsilon_{ij}f_{ij}+\epsilon_{ij}D_i\hat
A_j+\epsilon_{ij}{\theta_1\over 2}\{\hat A_i,\hat A_j
\}=Bq\nu_1\nu_2 ,\ee where \be\label{46} D_i\hat A_j=
{\partial\over\partial z_i}\hat A_j+\theta_2 \{a_i,\hat A_j \} .\ee

Again, as in \cite{susskind}, the above formula as well as the
discreteness of quasiparticles implies the full noncommutative
assumption. Together with the results in Section 2,
the full system is described by \bea L&=&L_1+L_2+L_{int} \label{47}\\
L_1&=&{1 \over 4 \pi \nu_1} \int d^2y\epsilon_{\mu \nu \rho} \left(
A_{\mu}\ast \partial_{\nu} A_{\rho} +{2i \over 3} A_{\mu} \ast
A_{\nu}
\ast A_{\rho} \right) \label{48}\\
L_2 &=& -{q \over 4 \pi \nu_2} \int d^2z\epsilon_{\mu \nu \rho}
\left( a_{\mu}\star
\partial_{\nu} a_{\rho} +{2i \over 3} a_{\mu} \star a_{\nu}
\star a_{\rho} \right) \label{49}\\
L_{int}&=&{q\over 2\pi}\int d^2z \left(\epsilon_{ij}\left(\hat
A_i+{\theta_1\over 2}\epsilon_{lk} \hat A_l \hat \ast{\partial \hat
A_k\over\partial y_i}\right) \star f_{0j}-2\pi\ro'\hat
A_0+Bq\nu_1\nu_2a_0\right) \label{50} ,\eea where \bea f(y)\ast
g(y)&=&\exp\left({i\over 2}
\theta_1\epsilon_{ij}{\partial\over\partial
\xi^i}{\partial\over\partial
\zeta^j}\right)f(y+\xi)g(y+\zeta)|_{\xi=\zeta=0} \label{51}\\
f(z)\star g(z)&=&\exp\left({i\over 2}
\theta_2\epsilon_{ij}{\partial\over\partial
\xi^i}{\partial\over\partial
\zeta^j}\right)f(z+\xi)g(z+\zeta)|_{\xi=\zeta=0} \label{52}\\
f(y) \hat \ast g(y)&=&\sum_{n=0}^{\infty}{1\over
(n+1)!}\left({i\over 2} \theta_1\epsilon_{ij}{\partial\over\partial
\xi^i}{\partial\over\partial \zeta^j}\right)^n
f(y+\xi)g(y+\zeta)|_{\xi=\zeta=0} \label{53} .\eea Taking the
$\delta A_0$ and $\delta a_0$ variations of (\ref{47}) to give the
constraints, the leading terms are \bea {1\over
4\pi\nu_1}\epsilon_{ij} F_{ij}&=&q\ro'\left(1+{1\over
4\pi\ro'}\epsilon_{ij}f_{ij}\right)^{-1} \label{54}\\
\left(1+{1\over 4\pi\ro'}\epsilon_{ij}f_{ij}\right)
\left(B+{1\over2}\epsilon_{ij}\hat F_{ij}\right)
&=&B\left(1+q\nu_1\nu_2\right) .\label{55}\eea

Substituting the constraints back into (\ref{50}), we find
Lagrangian becomes \bea\label{56} L&=&{1 \over 4 \pi \nu_1} \int
d^2y\epsilon_{\mu \nu \rho} \left( A_{\mu}\ast
\partial_{\nu} A_{\rho} +{2i \over 3} A_{\mu} \ast A_{\nu}
\ast A_{\rho} \right) \cr &-& {1\over 4 \pi \tilde\nu_2} \int
d^2z\epsilon_{\mu \nu \rho} \left( a_{\mu}\star
\partial_{\nu} a_{\rho} +{2i \over 3} a_{\mu} \star a_{\nu}
\star a_{\rho}\right)- \int d^2z q\ro'\hat A_0 .\eea If we separate
out the matrix model of quasiparticle, the coefficient ${1/ 4 \pi
\tilde\nu_2}$ would relate to the correct statistics of
quasiparticles as we hope.

There are some subtleties for the corresponding matrix model since
the two kind of non-commutativity in our case. In a naive
construction, we can replace the fields in (\ref{41}) by matrices
straightforwardly and arrive at \bea\label{57} L&=&q
Tr\{\left(-{B\over 2}\epsilon_{ij} y_j+ A_i+{i\over 2}\epsilon_{jk}
A_j [A_k, \epsilon_{il}y_l]\right)\left(\dot
y_i+i[y_i,a_0]\right)\cr &&+{B\over
2\pi\ro'}\left(1+q\nu_1\nu_2\right)a_0-A_0\} ,\eea where
\be\label{58} A_i=2\pi\ro\epsilon_{ji} (x_j-y_j) .\ee When the
number of electron and quasiparticle are $K$ and $\nu_1 K$
respectively, the matrices $x_i$ are $K\times K$ while $y_i$ are
$\nu_1K\times \nu_1 K$. We may simply take $y_i\otimes I_{p_1\times
p_1}$ as $y_i$ in the above expression.

\section {Comparison with K-matrix theory}

Our result is related to the effective Wen and Zee's K-matrix
Chern-Simons theory introduced in \cite{wen}. If we convert the
Lagrangian (\ref{49}) and (\ref{50}) to $y$-plane by using the
relation (\ref{22}). Although the full theory looks rather
complicated after this operation, the leading term is rather simple
and interesting. It is \bea\label{59} L&=&{1 \over 4 \pi \nu_1} \int
d^2y\epsilon_{\mu \nu \rho} A_{\mu}
\partial_{\nu} A_{\rho}
 \cr &-&{q \over 4 \pi \nu_2} \int d^2y\epsilon_{\mu
\nu \rho} a_{\mu}
\partial_{\nu} a_{\rho} +{q\over 2\pi}\int
d^2y\left(\epsilon_{\mu \nu
\rho}A_{\mu}\partial_{\nu}a_{\rho}-2\pi\ro' A_0\right)\cr&+&{q \over
2 \pi } \int d^2yBq\nu_1\nu_2a_0(1-\epsilon_{ij}\partial_{i}a_{j})
.\eea The result is analogy with the action for one generation
K-matrix Chern-Simons theory. Especially, we find the coefficients
in our result is fit with those in \cite{wen} which is used to
describe the K-matrix as well as topological order. The main
differences in between is the meaning of the gauge fields. The gauge
fields are the coordinates of the particle in our case while they
are related to the current of the particle in K-matrix Chern-Simons
theory. We employed the Lagrangian picture of fluid and K-matrix
Chern-Simons theory is explained in the Eulerian picture.

Another important thing is that the Chern-Simons description of edge
excitation in quantum Hall state introduced in \cite{wen} would
receive a natural explanation in our frame work. To show this point,
let us take the simplest case where there is no quasiparticle
excitation as an example. Considering the system in \cite{wen}, an
extra electric field $E$ is introduced to confine the quantum Hall
fluid below the $x$-axes. The Lagrangian of single particle becomes
\be\label{60} L= \ro \int d^2y \left({e{\mathcal{B}} \over 2}\epa
\dot{x}_a x_b-eEx_2\right) .\ee The particles will get a extra
velocity from the electric field. Thus instead of (\ref{7}), we
would expand around the vacuum configuration as \be\label{61} x_i =
y_i +{\ep E_j\over\mathcal{B}} t+\ep {A_j \over 2\pi \ro} .\ee Then
the resulting $y$-plane Chern-Simons theory remains the same formula
as (\ref{10})\be\label{62} L= {e{\mathcal{B}} \over 8\pi^2 \ro} \int
d^2y \epa \dot{A}_a A_b ,\ee which is the standard $y$-plane
Chern-Simons Lagrangian in the temporal gauge. When we discuss the
physics on the $x$-plane, the transformation \be\label{63} x_i = y_i
+{\ep E_j\over\mathcal{B}} t ,\ee should be performed. That is just
the transformation for physical gauge choice in \cite{wen}. In
presence of a boundary of the system, the area preserving
diffeomorphism gauge symmetry is restricted naturally to those which
leave the boundary invariant as demanded in \cite{wen}. Especially,
our approach automatically gives out the velocity of edge excitation
$v={E\over\mathcal{B}}$ without comparison to other computation
which is needed in \cite{wen}.

From the above evidences, we conclude that the K-matrix Chern-Simons
theory could be regarded as certain commutative limit of our theory.
The main differences are the meaning of gauge fields as well as the
additional terms coming from non-commutativity. These differences
offer reasonable candidates for the experimental deviation \cite{ex}
to the results of tunneling exponent predicted by K-matrix theory.
We will discuss this subject in a subsequent work.

\section {Conclusion}

In this paper we have reviewed the interesting way of construct
non-commutative Chern-Simons description for FQHE with $\nu=1/ p$ in
\cite{susskind}. It is based on the area preserving diffeomorphisms
of continuous electron indexes under fluid approximation. We have
applied the same method on the quasiparticles and reformulated the
hierarchy construction of FQHE suggested in \cite{haldane}. The
crucial point is that the coordinates of quasiparticle becomes the
new gauge field on $z$-plane, where $z$-plane comes from a continuum
description of the quasiparticle indexes rather than a simple
coordinate choice. It is different from the case that one directly
adds another kind of particles on $x$-plane which is a direct
superposition of two (or more) independent Laughlin droplets. We
have also shown that the interaction between quasiparticles implies
the correct statistics of quasiparticles. The full non-commutative
Chern-Simons theory is also built. However, the corresponding matrix
model remains unclear because the coupling of the two kinds
non-commutativity particles living on different plane. Also, the
case is different in superposition of two independent Laughlin
droplets where no such coupling exists. We have offered some naive
proposal on the matrix model.  Our work is closely related to the K
matrix theory in \cite{wen}. We have provided an element derivation
of the Chern-Simons theory for hierarchy description of quantum Hall
state. The K-matrix Chern-Simons theory could be regarded as its
commutative limit. On some related discussions of noncommutative
Chern-Simons description of hierarchies FQH states, see
\cite{Saidi}\cite{Eliashvili}.

We have mentioned in Section 1 that there is another approach to
FQHE with generic filling fraction, the composite fermion
assumption. It is argued in \cite{Bergholtz} that the hierarchy
construction and composite fermions may be complementary views of
the same phenomena rather than mutually excluding descriptions. Some
researches such as \cite{Greiter} \cite{fradkin} have tried to
relate the Chern-Simons actions in \cite{wen} and those in the
composite fermion approach \cite{lopez}. Since the composite fermion
approach is favored by the experiment results, we hope to relate the
two sides under Susskind's framework in our future works.

\section{Acknowledgments}

This work is supported by the grants from the NSF of China with
Grant No: 10588503, 10535060, the grant from 973 Program of China
with grant No: 2007CB815401 and the Pujiang Talent Project of the
Shanghai Science and Technology Committee under Grant Numbers
06PJ14114.

\end{document}